# Supporting dry eye diagnosis with a new method for non-invasive tear film quality assessment

Clara Llorens-Quintana[1], MSc, Dorota Szczesna-Iskander[2], PhD

and D. Robert Iskander[1], PhD, DSc

[1]Department of Biomedical Engineering, Wroclaw University of Science and Technology, Wroclaw, Poland

[2]Department of Optics and Photonics, Wroclaw University of Science Technology, Wroclaw Poland

Corresponding author:

Clara Llorens-Quintana

Department of Biomedical Engineering,

Wroclaw University of Science and Technology

Wybrzeze Wyspianskiego 27, 50-370 Wroclaw, Poland

E-mail: clara.llorens.quintana@pwr.edu.pl

Financial interest of the authors: none

**Abstract**

**Significance:** Non-invasive high speed videokeratoscopy equipped with specific software has shown potential for assessing the homeostasis of tear film, providing clinicians with a fast and consistent tool for supporting dry eye diagnosis and management.

**Purpose:** To evaluate the efficacy of a recently proposed method for characterizing tear film dynamics using non-invasive high speed videokeratoscopy in assessing the loss of homeostasis of tear film.

**Methods:** Thirty subjects, from a retrospective study, of which 11 were classified as dry eye and 19 as normal, were included. High speed videokeratoscopy measurements were performed using E300 videokeratoscope (Medmont Pty., Ltd, Melbourne, Australia). Raw data was analyzed using a recently proposed method to estimate the dynamics of the tear film, based on a fractal dimension approach. This method provides three time-varying indicators related to the regularity of the reflected rings: Tear Film Surface Quality (TFSQ) indicator, Breaks Feature Indicator (BFI) and Distortions Feature Indicator (DFI). From each indicator five parameters were extracted and analyzed, including non-invasive break up time, mean value of the indicator in the stability phase, mean value of the indicator in the whole inter blink interval, mean value of the indicator in the levelling phase and the general trend of the time series. Receiver Operating Characteristic were used to determine the sensitivity and specificity of each parameter in dry eye detection.

**Results:** The best discrimination performance between dry eye and normal subjects was achieved with the BFI non-invasive break up time parameter, with an area under the curve of 0.85. For a cut off value of 10 s the sensitivity was 100% and the specificity 84%.

**Conclusions:** The analyzed method improves the assessment of tear film homeostasis in comparison to previous high speed videokeratoscopy methods showing higher potential in assisting dry eye diagnosis.

**INTRODUCTION**

Dry eye disease is a frequently reported pathological condition of the ocular surface with an index of prevalence that ranges from 5% to 50%, depending on the demographics and the diagnostic criteria used.[1] This wide variability on dry eye disease epidemiology could be attributed partly due to the lack of a standardized definition and classification system. The Definition and Classification Subcommittee of the International Dry Eye Workshop II (DEWS II, 2017), with the goal of creating an evidence-based classification, defined dry eye disease as a "multifactorial disease of the ocular surface characterized by a loss of homeostasis of the tear film, and accompanied by ocular symptoms, in which tear film instability and hyperosmolarity, ocular surface inflammation and damage, and neurosensory abnormalities play etiological roles".[2]

Tear film instability has been considered as one of the core mechanisms that causes dry eye disease.[3] Consequently, the stability of the tear film lipid layer,[4] assessed by non-invasively measured tear film break-up time, has been recommended, along tear osmolarity and ocular surface staining, as one of the essential biomarkers of loss of tear homeostasis in dry eye diagnosis.[2,5] Etiologically, dry eye disease can be classified in two major groups: evaporative dry eye and aqueous-deficient dry eye; in both conditions the stability of the tear film lipid layer is compromised.

Although the assessment of the stability of the tear film is one of the main tasks supporting dry eye diagnosis, there is no gold-standard diagnostic tool or standardized clinical protocol available.[6] Also, in the presence of a large number of tools and techniques, it is not evident, particularly in a clinical setting, which of them is the most appropriate.[7,8,9,10,11,12] Hence, accurately assessing and monitoring the stability of tear film is a difficult task.[13]

Tear film stability can be assessed with the measurement of the tear film break-up time.[14] Traditionally it is measured by instilling a drop of fluorescein into the eye and determining the time until the appearance of the first dark growing spot on the tear film with the aid of a biomicroscope equipped with a yellow filter (such as Wratten12). The fluorescein break-up time is invasive, subjective and has shown lack of reliability and repeatability. Likewise, the agreement between fluorescein break-up time and other clinical measures of dry eye is weak.[15,16] It has been claimed that the preferred technique to assess the stability of the tear film should be non-invasive, quantitative and objective.[5,6]

For this reason, in the last years the development of non-invasive automatic measurement systems to assess the stability of the tear film in a more natural state has gained importance.[8,17,18,19,20] In particular, the techniques that assess the quality of the tear film by the observance of morphological changes in the specular reflection of a grid pattern projected on the cornea have shown to be promising.[21,22] One of those techniques is the high speed videokeratoscopy, which could be viewed as an extension of the traditional static videokeratoscope.[10] The videokeratoscope is a Placido disk topographer that projects a set of concentric rings onto the ocular surface. The regularity of the reflected pattern will depend, first of all, on the stability of the tear lipid layer. In high speed videokeratoscopy, continuous captures of the reflected images from the cornea are recorded, providing dynamic information of changes in the tear film along the time between blinks.[21] High speed videokeratoscopy is an accessible tool for clinicians that it is easy to use. The development of specific software to analyze the recorded videos in an objective and automated fashion provides an added value to standard videokeratoscopy when assessing the tear film stability.

Different image processing techniques have been proposed to analyze high speed videokeratoscopy recordings based on the analysis of the raw images provided by the instrument and nowadays several commercially available videokeratoscopes already incorporate an automated tear film analysis function. [8,23,24,25] However, although the performance of these automated methods has been evaluated in dry eye subjects, there are only few studies that have reported the sensitivity and specificity of these techniques to diagnose dry eye.[26,27,28,29] Also the repeatability and agreement of the automated methods have been questioned.[30,31,32]

Recently we have proposed a novel, automated and objective technique that analyses the recordings of high speed videokeratoscopy to obtain an estimator of tear film surface quality (TFSQ) and non-invasive break-up time.[33] In this technique, the post-blink dynamics of the tear film are derived from a textural analysis of the videokeratoscopy recordings by means of computing the fractal dimension of the images. This method has been tested for subjects with healthy tear film, showing its utility in characterizing three different phases of tear film dynamics (i.e., levelling, stability and evaporation, described previously by Braun et al.[34]) and estimating the non-invasive break up time. Also, it was shown that the method based on fractal dimension provides estimates of tear film dynamics closely corresponding to the observable, by clinicians, deformations of Placido rings during an inter-blink interval.

The purpose of this study is to evaluate the proposed method of high speed videokeratoscopy image analysis in cohorts of normal and dry eye subjects and test its efficacy for supporting dry eye diagnosis.

**METHODS**

*Subjects and data acquisition*

Videokeratoscopy recordings from a previous study[27] were analyzed with the algorithm proposed earlier[33] to evaluate its performance in the assessment of tear homeostasis. That algorithm is based on textural analysis of Placido disk pattern. Data of right eyes of 30 subjects was used (20 females and 10 males). Subjects were enrolled in the study voluntarily after informed consent was given. The study was approved by the Queensland University of Technology research ethics committee and followed the tenets of the Declaration of Helsinki. An experienced clinician performed the clinical assessment of dry eye signs and symptoms. The tests included medical history, McMonnies questionnaire,[35] slit lamp examination, phenol red thread test of tear volume and fluorescein tear film break-up time. Dry eye was diagnosed if the three following conditions were met: McMonnies test score > 14, fluorescein tear film break-up time < 10 s and corneal or conjunctival staining score > 3. Note that the study preceded the DEWS II report. A detailed explanation of the clinical protocol and the clinical values for each group can be found in the work of Szczesna et al.[27] According to this classification the cohort was divided into 11 dry eye and 19 normal subjects.

Another masked clinician performed high speed videokeratoscopy measurements with E300 videokeratoscope (Medmont Pty., Ltd, Melbourne, Australia) with a sampling frequency of 25 frames per second. Measurements were taken in suppressed blinking conditions. For that, patients were asked to look at the fixation target of the instrument, blink gently a few times and maintain their eyes open as long as they could for a maximum time of 30 seconds. Three measurements per eye were taken with a 3-minute break between them. All the measurements were performed at approximately the same time of day in a room with monitored temperature and humidity. Measurements were performed from less to more invasive in order to avoid the disturbance of the tear film physiology in the subsequent tests.

*High speed videokeratoscopy analysis*

High speed videokeratoscopy recordings were stored and analyzed offline using a Matlab-based custom written algorithm to estimate the dynamics of the tear film. In the analysis, the texture (regularity) of the reflected Placido disk pattern is used to determine the stability of the tear film, using a fractal dimension approach.[33] An example of the appearance of breaks and distortions of the Placido disk reflection in a single high speed videokeratoscopy frame is shown in Figure 1.

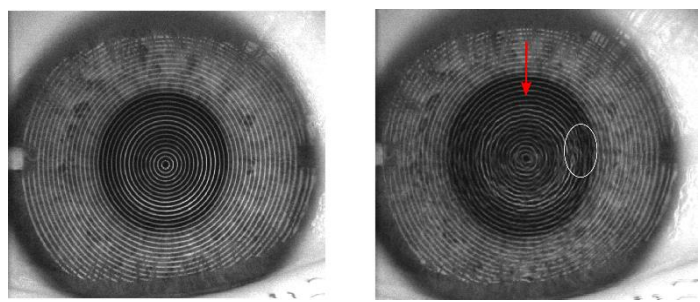

**Figure 1**. Appearance of the Placido disk pattern reflected on the ocular surface. Left: regular reflected rings from a stable tear film. Right: distorted reflected rings from a destabilized tear film. Red arrow indicates uneven rings which would contribute to the distortions feature indicator and white ellipse indicates broken rings which would contribute to the breaks feature indicator.

The fractal dimension is a measure of texture roughness. High values of fractal dimension are related to more irregular structures while values below one are related to incomplete structures. The fractal dimension is highly correlated with human perception of texture and relatively insensitive to changes in image intensity and scaling.[36] In previous work, it has been demonstrated that fractal dimension based assessment of high speed videokeratoscopy recordings is directly related to local morphological changes of the reflected pattern, thereby directly related to the regularity of the tear film. Briefly, the algorithm detects the inter-blink interval where the analysis is performed and uses an image block processing approach to compute the local fractal dimension of the images (i.e., the whole image is divided into small blocks and the fractal dimension of each sub-image is computed). Blocks are considered to have regular rings when the fractal dimension value is inside the interval of 1±0.18. From this analysis three time-series indicators, for each detected inter-blink interval, are obtained and used to describe the dynamics of the tear film:

a) Breaks Feature Indicator (BFI): this index is related to incomplete rings. For its calculation all the blocks of each image having a fractal dimension value below to the established interval are summed and the inverse of this summation is equal to BFI. Higher BFI values correspond to a greater number of incomplete rings (e.g., broken rings, example shown in Figure 2A).

b) Distortions Feature Indicator (DFI): this index is related to uneven rings. For its calculation all the blocks of each image having a fractal dimension value above the established interval are summed and this value corresponds to DFI. Higher DFI values correspond to a more irregular Placido disk pattern (example shown in Figure 2B).

c) Tear Film Surface Quality (TFSQ) indicator: this general index is directly related to the stability of the tear film. It is composed by the weighted combination of BFI and DFI. The higher the TFSQ the less regular the tear film is (example shown in Figure 2C).

A detailed explanation about how the different indicators are computed and the determination of the limits can be found in the previous work[33].

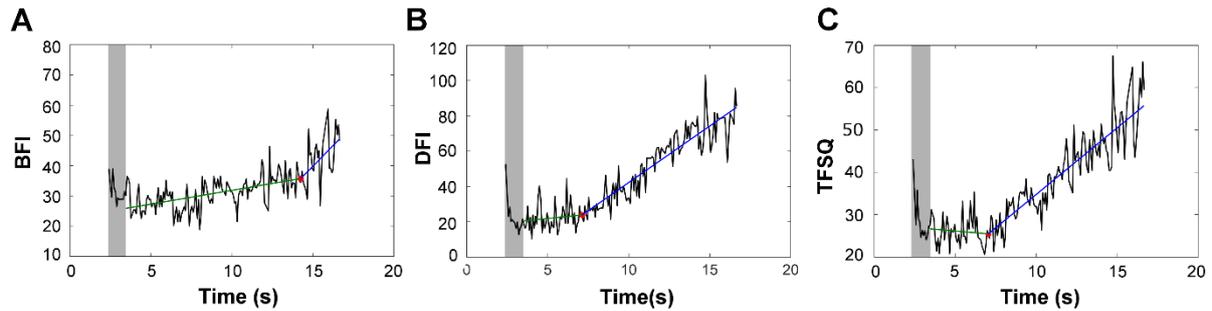

**Figure 2**. Example of the time series of the three considered indicators and the corresponding fittings of a bilinear model: A–BFI, B–DFI, and C–TFSQ index. All three indicators are estimated from a high speed videokeratoscopy recording of the same subject. The phases of tear levelling (gray shadowed area), stability (green line), deterioration (blue line) and the estimated break-up point (red dot) are demarcated.
BFI – Breaks Feature Indicator; DFI – Distortions Feature Indicator; TFSQ – Tear Film Surface Quality.

*Statistical analysis and data processing*

For the three repeated measurements of the same eye, the median values were considered for the analysis. Data corresponding to the first second after a blink was removed to avoid the possible effect of the initial phase of the tear film dynamics (known as the levelling phase). The Shapiro-Wilk test revealed that the data was not normally distributed ($P < .05$), accordingly, differences between medians groups were analyzed with the Mann-Whitney test for independent samples. A value of $P < .05$ was considered statistically significant.

To estimate the non-invasive break up time, the remaining raw time series data (omitting the first second) of each one of the three dynamic indicators (i.e., TFSQ, BFI and DFI) was fitted with three different functions: a linear, a constrained bilinear and a constrained linear-polynomial function. The bilinear and linear-polynomial functions are two-section functions. The former is comprised of two linear sections (see Figure 3C) whereas the second is comprised of a linear section followed by a second degree polynomial section (see Figure 3D). In both those cases, a constrain is made so that the point where the first section ends must correspond to the point where the second section starts (see red points in Figure 3C and 3D, this is the constraining point) in order to avoid discontinuity between the sections in the model.

Representative examples of the three fitting types can be seen in Figure 3. The suitability of the fittings was assessed by computing Pearson's correlation coefficient ($r^2$), between the raw data and the fitting, and testing the null hypothesis of equality in $r^2$ with the Fisher test. If the null hypothesis was not rejected ($P > .05$), the linear function was chosen, otherwise, the more appropriate constrained function (bilinear or linear-polynomial function) was fitted taking the constraining point as the estimated non-invasive break up time.

The three different models (linear, constrained bilinear and constrained linear-polynomial) were considered to precisely estimate the non-invasive break up time under the changing characteristics of the individual tear film dynamics.[37,38]

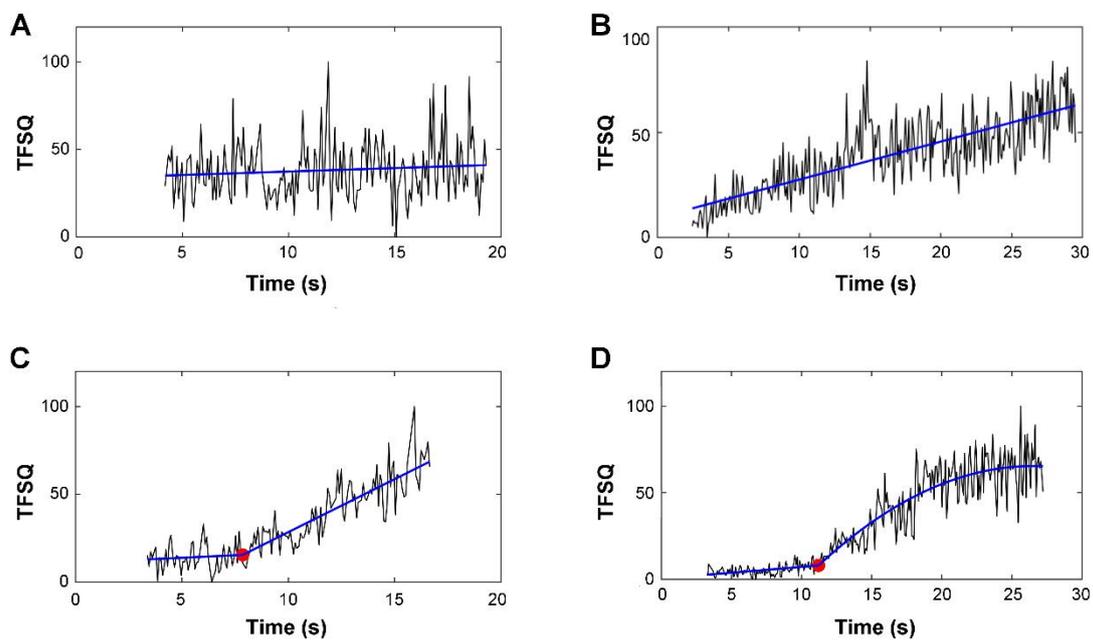

**Figure 3.** Representative examples of the different possible fittings. A–Linear fitting with neutral trend, NIBUT = inter-blink interval time. B–Linear fitting with a positive trend, NIBUT = 0. C–Bilinear fitting, the red dot demarcates the estimated NIBUT. D–Linear-polynomial fitting, the red dot demarcates the estimated NIBUT. NIBUT – non-invasive break up time.

In the case that a linear function was fitted, the estimated non-invasive break up time was set to either 0 s, if the slope of the linear function was greater than 10 degrees (see Figure 3B), or to the duration of the inter-blink interval if the slope of the linear function was 10 degrees or less (no non-invasive break up time was observed during the duration of the inter-blink interval, see Figure 3A).

In addition to the non-invasive break up time, other parameters of each indicator (TFSQ, BFI and DFI) were extracted and analyzed. These were:

- The mean value of the indicator along the stability phase (mean stability phase).
- The mean value of the indicator along all the considered inter-blink interval omitting the first second after the blink (mean inter-blink interval value).
- The general trend of the indicator (slope).

Mean values of the indicators in the different phases of the inter-blink interval are related to the overall quality of the tear film in the given phase. They are computed in order to clarify if the difference between dry eye and normal subjects only lies in the non-invasive break up time or if, on the contrary, there is also a difference in the quality of the tear film even when it is stable. On the other hand, the general trend of each indicator determines the speed with which the tear film evaporates. This metric is computed in order to explain if once the tear film has been destabilized the evaporation is quicker for dry eye subjects than that observed in normal subjects.

The Receiver Operating Characteristic (ROC) curves were used to determine the sensitivity (true positive rate) and specificity (1−false positive rate) of the tested algorithm for dry eye diagnosis. To create ROC curves the probability density function for normal and dry eye subjects of each considered parameter was computed using a kernel density estimator with an Epanechnikov window.[39] From each ROC curve, in addition to the sensitivity and specificity, other statistical parameters that provide information about the discrimination performance of the method were extracted.[40] These were:

- Area under the ROC curve (AUC), computed using trapezoidal numerical integration. It is bounded between 0 and 1. The closer to 1 the better the performance of a detector.
- Cut-off value that optimizes the discrimination between normal and dry eye subjects, determined as the point for which the distance between the ROC and the diagonal is maximum.

- Youden's index,[41] defined as: $\gamma = sensitivity + specificity - 1$, so the closer to 1 the better the performance of a detector.

- Discriminant power,[42] defined as: $DP = \frac{\sqrt{3}}{\pi}(\log(\frac{sensitivity}{1-sensitivity}) + \log(\frac{specificity}{1-specificity}))$, where $DP < 1$ means poor discrimination, $1 < DP < 2$ means limited discrimination and $DP < 3$ means fair discrimination and values above 3 are considered as good discrimination.

**RESULTS**

The mean ± standard deviation of the inter-blink interval duration for normal and dry eye subjects was $24.2 \pm 6.3$ s and $14.9 \pm 8.6$ s, respectively, and the difference was statistically significant ($P < 0.0001$). As a result of the fitting the mean ± standard deviation TFSQ non-invasive break up time was $16.4 \pm 8.3$ s for normal subjects and $8.6 \pm 2.9$ s for dry eye subjects.

Table 1 shows the descriptive statistics (median and interquartile ranges) for all the considered parameters and the *P values* for the Mann-Whitney test between normal and dry eye subjects. With the exception of the mean stability and the mean inter-blink interval values for DFI, all the parameters showed inferior tear film quality characteristics for the dry eye subjects. Statistically significant differences were found for the non-invasive break up time assessed by all three indicators (i.e., TFSQ, BFI and DFI), the slope of TFSQ and DFI and the mean stability phase value of BFI.

ROC curves were computed for all the parameters of each indicator. Table 2 summarizes the diagnostic power for those parameters with the best discriminative performance. The BFI non-invasive break up time showed to be the most powerful indicator in differentiating between normal and dry eye subjects, the ROC of this parameter is shown in Figure 4.

Table 1

Median values and interquartile ranges for the parameters computed for each tear film stability indicator for normal and dry eye subjects.

| Indicator | Parameter | Normal subjects | | DE subjects | | P |
|---|---|---|---|---|---|---|
| | | Median | IQ | Median | IQ | |
| TFSQ index | NIBUT (s) | 14.2 | 13.7 | 8.7 | 3.0 | .02* |
| | Mean stability phase vale | 43.6 | 9.7 | 51.4 | 10.5 | .21 |
| | Mean inter-blink interval value | 49.093 | 12.4 | 51.5 | 12.8 | .44 |
| | Slope (deg.) | 19.4 | 29.1 | 65.7 | 44.5 | .004* |
| BFI | NIBUT (s) | 21.4 | 17.3 | 7.8 | 3.5 | .001* |
| | Mean stability phase value | 39.7 | 6.1 | 44.6 | 9.6 | .049* |
| | Mean inter-blink interval value | 40.0 | 8.1 | 44.6 | 10.2 | .19 |
| | Slope (deg.) | 28.0 | 20.8 | 31.8 | 55.9 | .21 |
| DFI | NIBUT (s) | 15.1 | 9.2 | 8.7 | 6.1 | <.0001* |
| | Mean stability phase value | 53.0 | 43.8 | 42.6 | 27.6 | .21 |
| | Mean inter-blink interval value | 64.6 | 43.5 | 47.3 | 45.0 | .87 |
| | Slope (deg.) | 40.8 | 51.3 | 74.4 | 30.7 | .047* |

*$P < 0.05$
TFSQ index: Tear Film Surface Quality index
BFI: Breaks Feature Indicator
DFI: Distortions Feature Indicator
NIBUT: Non-invasive break up time

Table 2

Statistical parameters that determine the efficacy in dry eye diagnosis for the three considered parameters.

| Parameter | Sensitivity | Specificity | AUC | Cut-off value | γ | DP |
|---|---|---|---|---|---|---|
| BFI-NIBUT | 100% | 84% | 0.85 | 10 s | 0.84 | 4.71 |
| Slope TFSQ | 100% | 48% | 0.73 | 25 deg. | 0.48 | 3.76 |
| DFI-NIBUT | 100% | 56% | 0.62 | 11.1 s | 0.56 | 3.94 |

AUC: Area Under the Curve
γ: Youlden's Index
DP: Discriminant Power
BFI-NIBUT: Non-Invasive Break Up Time for Breaks Feature Indicator
TFSQ: Tear Film Surface Quality Index
DFI-NIBUT: Non-Invasive Break Up Time for Distortions Feature Indicator

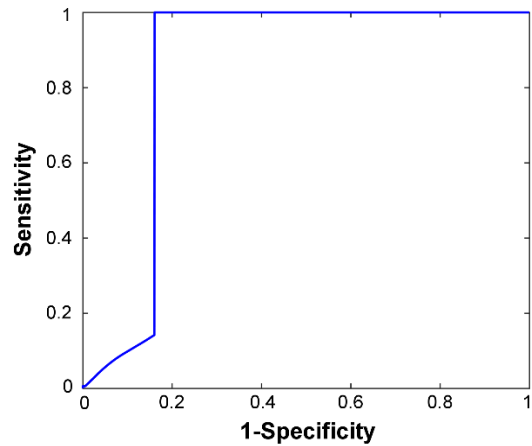

**Figure 4.** The ROC curve for the BFI (Breaks Feature Indicator) non-invasive break up time parameter.

For a cut off value of 10 s its sensitivity is 100% and the specificity 84% (subjects with BFI non-invasive break up time lower than 10 s are classified as dry eye and higher as normal). This means that, for a BFI non-invasive break up time > 10 s a subject could be classified as normal with certainty that there are no false negatives. Although this parameter is powerful enough to be a good classifier on its own, there is still a probability (16%) that a normal subject is misclassified as dry eye (false positive). Given that one of the characteristics of this method is that it provides multiple indicators, the rate of false positive can be decreased by performing a sequential analysis with another parameter. The proposed process is schematized in Figure 5; thereby for a subject to be classified as potentially having dry eye it needs to meet two conditions: a BFI non-invasive break up time < 10 s and a DFI non-invasive break up time < 11.1 s. Following this approach, in the representative cohort of this study three normal subjects had a BFI non-invasive break up time < 10 s, so they would be misclassified as dry eye if the BFI non-invasive break up time cut off value was taken into account as the only indicator. However, applying the proposed sequential analysis, the percentage of false positive would be reduced to 7.4%.

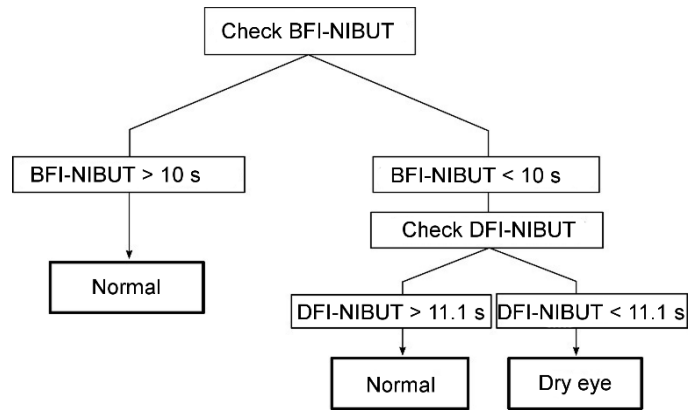

**Figure 5**. Proposed schema for a dry eye sequential analysis
BFI-NIBUT: Non-Invasive Break Up Time for Breaks Feature Indicator
DFI-NIBUT: Non-Invasive Break Up Time for Distortions Feature Indicator
DED: dry eye disease

**DISCUSSION**

Although dry eye is an ophthalmic disease affecting a large part of the population, a unified and reliable approach to its diagnosis is still missing.[13] Traditional diagnostic tests are invasive, qualitative and/or subjective, potentially leading to misdiagnosis and inaccurate treatment follow up.[16] Previous studies have demonstrated the utility in assessing tear film stability in dry eye diagnose[43,44] and the importance of using non-invasive and objective techniques due to the lack of repeatability and reproducibility of the traditional tests. [22,45,]

Non-invasive and objective methods that allow the assessment of tear stability are already commercially available and implemented in some videokeratoscopes (e.g., Oculus Keratograph, Medmont E300, and Tear Stability Analysis System). They are equipped with a specific software to estimate the non-invasive break up time based on the reflection of a pattern projected onto the ocular surface. However, dry eye diagnosis with such instruments is still challenging because, among other things, establishing the cut off values between healthy and unhealthy populations is difficult. Values of currently available metrics are continuous, not showing dichotomous behavior between dry eye and normal subjects.

In this study, the capability to diagnose dry eye of a recently proposed method, which analyses videokeratoscopy recordings, has been tested. This method provides different dynamic indicators related to the regularity of the reflected pattern in videokeratoscopy, thereby, related to the dynamics and stability of the tear film. The diagnostic ability of three indicators has been tested in order to arrive at the best detection criteria. The TFSQ non-invasive break up time is the parameter that is directly related to the first observed disturbance of the reflected rings (this is what a clinician would note as the non-invasive break up time), whereas BFI non-invasive break up time relates to completely broken rings and DFI non-invasive break up time to uneven rings.

The group mean TFSQ non-invasive break up time values for normal and dry eye patients were 16.4 s and 8.6 s, respectively. These values are slightly higher than those found with the fluorescein break-up time for that group: 14.3 s for normal subjects and 6.3 s for dry eye subjects.[27] This fact has already been widely reported since the stability of the tear film is affected by fluorescein.[46,43,47] Recently, Downie[29] has found longer non-invasive break up time values using the same type of videokeratoscope, but with different image analysis approach that is proprietary: 21.3 s for normal and 13.4 s for dry eye subjects. Hence, when comparing such results, it is important not only to consider the technique used, but also the approach followed to analyze the recorded sequences. In a study conducted by Abdelfattah et al.,[32] in which they used Oculus Keratograph (K5M), they did not found statistically significant differences between dry eye (mean non-invasive break up time of 8.2 s) and normal (mean non-invasive break up time of 6.7 s) populations, showing, in their case, the powerlessness to dry eye diagnosis. Results discordance may be influenced by the technical differences between E300 and K5M videokeratoscopes; K5M has lower number of mires, smaller corneal coverage and thicker rings which presumably can result in less precise detection of mire distortions.[48] In contrast, Hong et al.[29] did found differences using the K5M videokeratoscope, but lower values were reported for non-invasive break up time in normal subjects (4.3 s) and dry eye subjects (2.0 s). Many factors could have had a role in non-invasive break up time differences between the results of Hong et al. and the ones obtained in this study. First, the technical characteristics of the instrument. Second, the image analysis approach in K5M software is based on finding differences in the brightness of the data points on the mire rings, while in this study it is based on a morphological analysis of the deformation of the reflected Placido disk rings. And finally, the studied population was Asian, where higher prevalence of dry eye has been found.[49]

[Insert Table 3 here]

In terms of dry eye detection, the BFI non-invasive break up time showed the best discrimination performance between dry eye and normal subjects, with a sensitivity of 100% and a specificity of 84% for a cut off value of 10 s, and an AUC of 0.85. In Table 3, the results of the discrimination performance between dry eye and normal subjects from different studies are summarized. Results show an improvement on the discrimination performance using this approach over that achieved, with the same data set but using different image analysis approach, by Alonso-Caneiro et al.[26] and Szczesna et al.[27] Downie[28] also tested the diagnostic power of E300 using the machine's own software to estimate non-invasive break up time in dry eye and normal subjects. Although higher AUC and specificity were reported, the sensitivity was lower as well as the Youden's index and the discriminant power (see table 3).

Non-invasive examination of the tear film in natural blinking condition could be performed using the methodology reported earlier[33], however, the modeling of tear film indicator time series would require a different approach. To do so, the analysis should be focused on modeling the leveling phase.

Summarizing, the tested method to analyze E300 videokeratoscopy images, based on a fractal dimension approach, improves the detection performance in comparison to the previous approaches (i.e., image coherence and block feature). Although BFI non-invasive break up time has shown to be powerful enough to be a good dry eye detector, the discrimination performance can be improved by performing a consecutive analysis with a second indicator (DFI non-invasive break up time), following this approach if a subject has BFI-NIBUT > 10 s is classified as normal, but if it has BFI-NIBUT < 10 s, to be classified as a potential dry eye it has to meet the second condition of DFI-NIBUT < 11.1 s.

The implementation of the algorithm in high speed videokeratoscopy can be utilized in the clinical practice providing clinicians with a fast and consistent tool for supporting dry eye diagnosis and management that does not rely on subjective judgments. In addition, it can be used in the assessment of pre-lens tear film in contact lens wearers, supporting contact lens fitting[50].

It is evident that high speed videokeratoscopy based assessment of tear homeostasis for supporting dry eye diagnosis is dependent on different factors. First, it has to been taken into account that the criteria used to define dry eye subjects may vary between studies. Accordingly, those subjects that are close to the border line between normal and dry eye may be included in different groups depending on the criteria and dry eye definition used. Then, the results will depend on the technical characteristics of the instrument (e.g., number of rings, width of the rings or area of coverage), but also, for the same instrument, they will depend on the approach followed to analyze the images, since it is likely that different approaches are measuring different characteristics of tear film deformation. This evidences the need of finding a unified approach that can be applied to different instruments utilizing high speed videokeratoscopy technique, ensuring some consistency in the assessment of tear homeostasis.

**Acknowledgments**

This project has received funding from the European Union's Horizon 2020 research and innovation program under the Marie Skłodowska-Curie grant agreement No. 642760.